\documentclass[twocolumn,preprintnumbers,amsmath,amssymb]{revtex4}
\usepackage{graphicx}

\begin{document}

\title{Double injection, resonant-tunneling  recombination, and  current-voltage characteristics in
 double-graphene-layer structures} 
\author{M. Ryzhii,$^{1}$ V. Ryzhii,$^{2,3}$  T. Otsuji,$^{2}$ P. P. Maltsev,$^3$ 
V. G. Leiman,$^4$   N. Ryabova,$^{1,5}$ and  V. Mitin$^6$  } 
\affiliation{
$^1$ Department  of Computer Science and Engineering,
University of Aizu,
Aizu-Wakamatsu 965-8580, Japan\\
$^2$ Research Institute for Electrical Communication, Tohoku University, 
Sendai 980-8577, 
 Japan\\
$^3$ Institute of Ultra High Frequency Semiconductor Electronics, Russian Academy of Sciences, Moscow 111005, Russia\\
$^4$ Moscow Institute of Physics and Technology,
Dolgoprudny, Moscow Region 141700, Russia\\
$^5$ Center for Photonics and Infrared Engineering, Bauman Moscow State Technical University,
Moscow 105005, Russia\\
$^6$ Department of Electrical Engineering, University at Buffalo, Buffalo, New York 1460-1920, USA\\
}

\begin{abstract}
We evaluate  the effect of the recombination associated with interlayer transitions  in ungated and gated double-graphene-layer (GL) structures on the 
injection of electrons and holes. Using the proposed model,  we derive analytical expressions for  the spatial
distributions of the electron and hole  Fermi energies and the energy gap between the Dirac points in GLs as well as their dependences on the bias and gate voltages. The current-voltage characteristics are calculated as well.
The model is based on hydrodynamic equations for the electron and hole transport in GLs
under the self-consistent electric field. It is shown that in undoped double-GL structures with weak scattering of electrons and holes on disorder, the Fermi energies and the energy gap are virtually constant across the main portions of GLs, although their values strongly
depend on the voltages and recombination parameters. In contrast, the electron and hole scattering on disorder lead to substantial  nonuniformities. 
The resonant inter-GL tunneling enables N-shaped current-voltage characteristics provided that GLs are sufficiently short. The width of the current maxima is much larger than the broadening of the tunneling resonance. In the double-GL structures with relatively
long GLs the N-shaped characteristics transform into the Z-shaped characteristics.
The obtained results are in line with the experimental observations~\cite{1} and might be useful for design and optimization  of different devices based on double-GL structures, including field-effect transistors and terahertz lasers. 
\end{abstract}

\maketitle
\newpage
\section{Introduction}

Double-graphene-layer (double-GL) structures, which consist of two GLs separated by a thin  layer and independently connected with   the side contacts, 
were fabricated and studied recently~\cite{1,2,3,4}.
Such  structures have  drawn a 
considerable attention. This, in part, is  due to  potential applications of the double-GL structures in optically transparent transistor circuits~\cite{1,2,3},
high speed modulators of optical and terahertz (THz)  radiation~\cite{3,4,5,6},
THz detectors and frequency multipliers~\cite{7,8,9},  THz photomixers~\cite{10},  and  THz lasers~\cite{11}. 
Schematic views of the double-GL structures (both ungated and gated) and their band diagram under the bias voltage are shown in Fig.1.  Interesting features of devices based on independently contacted quantum wells, formed in
the standard heterostructures, were discussed some time ago~\cite{12}.
 Under the operation conditions of different double-GL-based devices,
the inter-GL transitions (tunneling or thermionic) tend to depopulate GLs. 
This can lead to a disruption of the device operation.
The refilling 
of GLs is associated with the injection of electrons  to one GL and holes another from
the pertinent contacts. Thus, the injection of electrons and holes in the double-GL structures requires a careful consideration.
In this paper, we develop a model for transport phenomena in the double-GL structures
shown ind Fig.~1(a) and 1(b)  employing   the hydrodynamic equations coupled with the Poisson equation for
 the self-consistent   electric potential~\cite{13,14}. 
 Similar problem, but related to single- and multiple-GL  structures with injection of both electrons and holes
 into the same GL from the opposite n- and p-contacts, was considered recently~\cite{13,14}.
However, the double-GL structures with independently contacted GLs are characterized by
important features associated with spatial separation of interacting   two-dimensional electron Gas (2DEG) and two-dimensional holes gas (2DHG). In particular,
 the existence of the energy gap between the Dirac points in GLs
and the resonant tunneling between GLs  should be addressed. 

The paper is organized as follows. In Sec.~II, we discuss the device model and write down the main equations which govern the transport and recombination of the injected electrons in the main parts GLs, where the latter are quasi-neutral. The role of the near-contact regions is accounted by the boundary condition set at the edges of the quasi-neutral regions~\cite{14}. Section~III deals with the calculations of the spatial distributions and voltage dependences of the Fermi energy, the energy gap, and electric potential assuming that the electron-hole scattering dominates  over the scattering on disorder. In Sec.~IV, we found how the disorder scattering affects the spatial distributions and voltage dependences. In Sec.~V, the current-voltage characteristics are calculated and discussed using the results of Sec.~IV.  In Sec.~VI, we draw the main conclusions. 

\section{Model and the pertinent equations}

We assume that GLs in the double-GL structures under consideration  are  ungated and undoped  [see Fig.~1(a)], or gated and
"electrically" doped. In the latter case, GLs are 
filled with 2DEG and 2DHG [see Fig.~1(b)] electrically induced by the gate voltages $\pm V_g/2$.
The application of the bias voltage $V$ between
the opposite edges of the ungated GLs [as shown in Fig.~1(a)] results in the formation
 of 2DEG and 2DHG in the pertinent GLs. In the gated double-GL structure shown in Fig.~1(b), the 2DEG and 2DHG densities are determined  by both the bias and gate voltages, $V$ and $V_g$.
 The double-GL structures under consideration are 
 particularly interesting for devices utilizing the resonant tunneling between GLs
and resonant tunneling assisted by the photon emission~\cite{11} 

As shown previously by numerical solutions of  the two-dimensional Poisson equation with realistic boundary conditions and  structural parameters ~\cite{14},
 in  sufficiently long  GL- structures (their length   $2L$ significantly exceeds the characteristic screening length $r_s = (\kappa\hbar^2v_W^2/4e^2T)$, i.e.,  the parameter
$Q =  2L/r_s   = (\pi\,e^2LT/6\kappa\hbar^2v_W^2)  \gg 1$, 
in the main part of GLs
the electron-hole plasma 
is quasi-neutral. 
Here $T$ is the temperature (in the energy units), $v_W \simeq 10^8$~cm/s is the characteristic velocity of electrons and holes in GLs, $\kappa$ is the dielectric constant, and $\hbar$ is the Planck constants. Indeed, if  $\kappa = 4$ and $2L = 10~\mu$m at the temperatures
$T = 300$~K, parameter $Q \simeq 600$. Thus the electron density in one GL,$\Sigma_e$,
is  with high accuracy equal to the hole density  $\Sigma_h$ in another GL:
$\Sigma_e = \Sigma_h = \Sigma$. Figure~2 schematically shows the Dirac points spatial positions (separated by the energy gap $\Delta$) and potential profiles in both GLs under the applied bias voltage $V$. 
This figure demonstrates qualitatively that the potential profiles being rather sharp near the contact regions are smooth in the main parts of GLs.

At sufficiently high bias and gate voltages,
the   2DEG and 2DHG density $\Sigma$ are large, so that the electron and hole energy distributions  
are well characterized by the Fermi functions with the quasi-Fermi  energies (counted from the Dirac points)
$\varepsilon_{Fe} =\varepsilon_{Fh} =\varepsilon_F$ and the common effective temperature $T$.
The latter is assumed to be equal to the lattice temperature. 

 The electron and hole density in the pertinent GLs $\Sigma$ and the energy gap between the Dirac points are related to each other as is 

\begin{equation}\label{eq1}
\Sigma = \Sigma_g + \frac{\kappa\Delta}{4\pi\,e^2d},
\end{equation}
where $\Sigma_g = \kappa\,V_g/8\pi\,eW_g$ is the density of 2DEG and 2DHG induced by the gate
voltages $\pm V_g/2$, 
  $d$ is the spacing between GLs and $W_g$ is the spacing between the GLs and gates [see Figs.~1(a) and 1(b)]
 Generally , the energy gap $\Delta$, i.e., the energy separation between the Dirac points in GLs, 
(which determines the built-in 
electric field $E = \Delta/ed$ 
in the inter-GL barrier),  as well as $\varepsilon_F$ and the electric potential
at the GL plane $\varphi $  are  functions of coordinate $x$ (the $x$-axis is directed in the Gl plane from the p- to n-contacts).

In the case of the 2DEG and 2DHG    strong degeneration ($\varepsilon_{F}\gg T$) , 

\begin{equation}\label{eq2}
 \varepsilon_F \simeq \hbar\,v_W\sqrt{\pi\Sigma},
\end{equation}
\begin{equation}\label{eq3}
 \Delta \simeq \frac{4e^2d}{\kappa\hbar^2\,v_W^2}(\varepsilon_F^2 - \varepsilon_{Fg}^2),
\end{equation}
where $ \varepsilon_{Fg} \simeq \hbar\,v_W\sqrt{\pi\Sigma_g} = \hbar\,v_W\sqrt{\kappa\,V_g/8eW_g}$.

In the absence of the inter-GL recombination, 2DEG and 2DHG are in equilibrium. In this case, 
Fermi energy in each GL and the energy gap are given by
\begin{equation}\label{4}
 \varepsilon_F^{eq} = \frac{e}{2}[\sqrt{(2VV_0 + V_0^2 + V_{bi}^2)} - V_0], 
\end{equation}
\begin{equation}\label{5}
 \Delta^{eq}  = e[V + V_0 -\sqrt{(2VV_0 + V_0^2 + V_{bi}^2)}],
\end{equation}
where $V_0 = \hbar^2v_W^2\kappa/2e^3d$ and  $V_{bi} = 2\varepsilon_{F_g}/e$.
At $V < V_{bi}$, $\Delta^{eq} < 0$, but at $V \geq V_{bi}$, $\Delta^{eq} \geq 0$. In particular,
when $\Sigma_g = 0$, i.e., $V_{bi} = 0$, $\Delta_{eq} \geq 0$ at all $V$.
Figure~3 shows the energy diagrams of a double-GL structure at $\Delta = 0$
and $\Delta > 0$.

High densities of 2DEG and 2DHG assume the validity of a hydrodynamic approach
for the description of  the electron and hole transport. We use the hydrodynamic model  presented  in Ref.~~\cite{14}.

\begin{figure}[t]
\center
{\includegraphics[width=8.5cm]{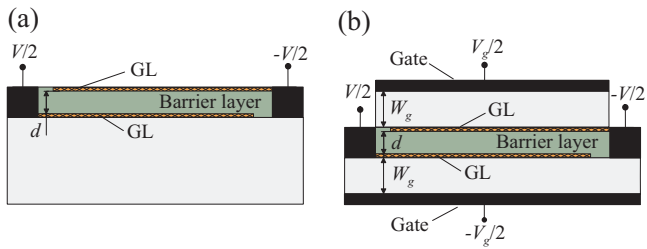}}
\caption{(Color online) Schematic view of (a) the cross-sections
of a double-GL structure and  (b) the cross-section of its version with the top and bottom metal gates.
}
\label{Fig1}
\end{figure}
\begin{figure}[t]
\center{\includegraphics[width=7.5cm]{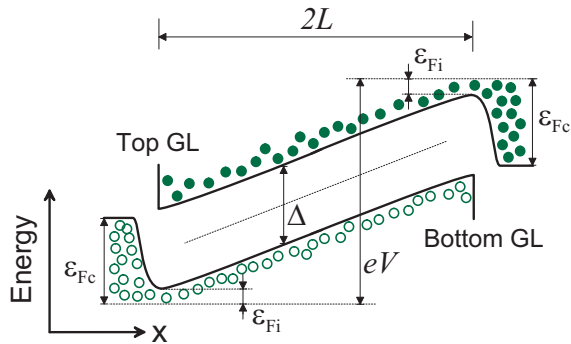}}
\caption{(Color online) Schematic view of spatial variations  of the Dirac points and potential profiles in top 
and   bottom GLs, respectively,  in a double-GL structure at $\Delta > 0$. Filled and open circles correspond to electrons in the conduction band of top GL and to holes in the valence band of bottom GL. 
}
\label{Fig2}
\end{figure}
\begin{figure}[t]
\center{\includegraphics[width=5.5cm]{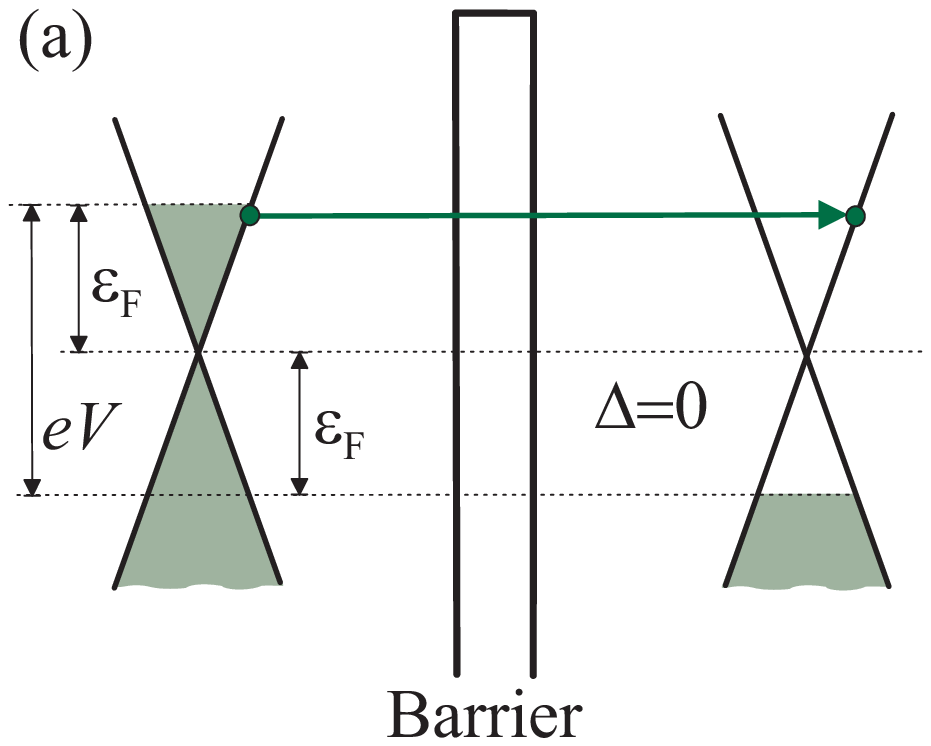}}
\center{\includegraphics[width=5.5cm]{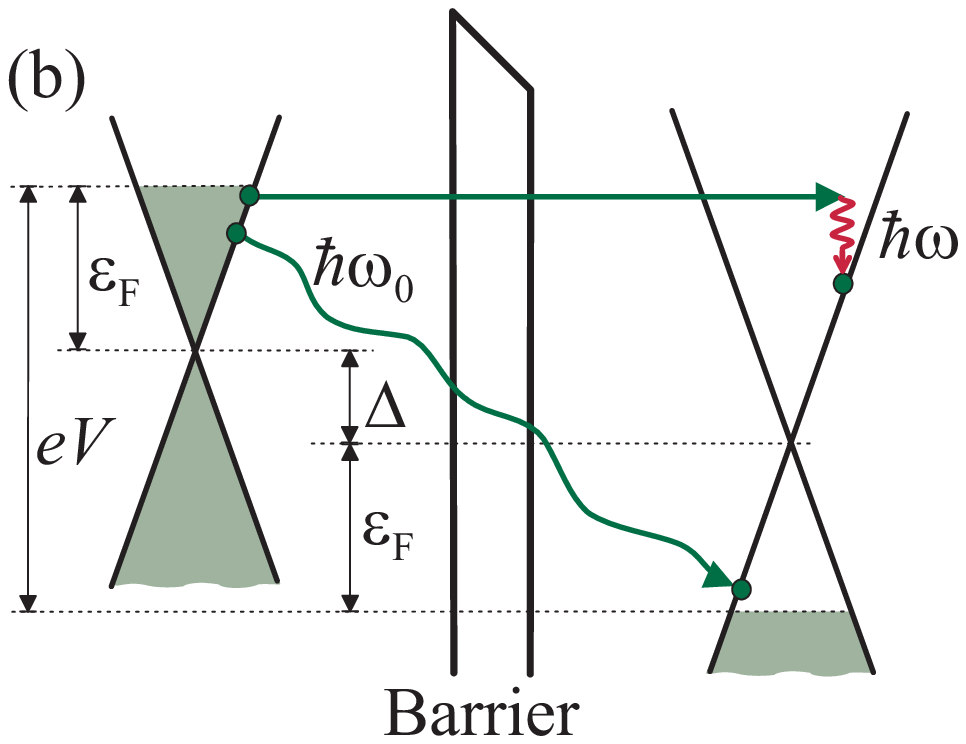}}
\caption{(Color online) Energy band diagrams 
of double-GL MGL structure (a) at $\Delta =  \Delta^{eq} = 0$, $eV < \hbar\omega_0$ and (b) at $\Delta 
= \Delta^{eq} > 0$, $eV > \hbar\omega_0$.
Arrow in (a)  indicates inter-GL resonant-tunneling transition (with conservation of electron energy and momentum), arrow in (b) correspond to transitions assisted by photon  and  optical phonon emission (with energies $\hbar\omega$ and $\hbar\omega_0$, respectively. 
}
\label{Fig3}
\end{figure}

The transport of electrons and holes is governed by the following system of hydrodynamic equations~\cite{14}:

\begin{equation}\label{6}
\frac{d \Sigma\,u_e}{d x} = -R, \qquad \frac{d \Sigma\, u_h}{d x} = -R.
\end{equation}

\begin{equation}\label{7}
\frac{1}{M}\frac{d(e\varphi_e - \varepsilon_{F})}{dx} = \nu\,u_e + \nu_{eh}(u_e - u_h),
\end{equation}

\begin{equation}\label{8}
-\frac{1}{M}\frac{d(e\varphi_h + \varepsilon_{F})}{dx} = \nu\,u_h + \nu_{eh}(u_h - u_e).
\end{equation}

\begin{equation}\label{9}
\Sigma = \frac{12\Sigma_T}{\pi^2}\int_0^{\infty}\frac{dyy}{[\exp(y - \varepsilon_{F}/T) +1]}.
\end{equation}
Here  $u_e$ and $u_h$ are the value of the  electron and hole 
hydrodynamic velocities  and $\varphi_e$ and $\varphi_h$ are the potentials of the top GL (filled with electrons) and the bottom  GL (filled with holes),  respectively,  $\nu$ is the  collision frequency of electrons and holes with impurities and acoustic phonons (with the  short-range disorder), $\nu_{eh}$ is their collision frequency  with each other,
$M$ is the fictitious mass, which at the Fermi energies of the same order of magnitude as
the temperature can be considered as a constant, and $\Sigma_T = \pi\,T^2/6\hbar^2v_W^2$.

The recombination of the major carriers injected to the  degenerate 2DEG and 2DHG in the double-GL structures under consideration is primarily associated with the inter-GL transitions. The rate of these processes depends of the Fermi energy (density of 2DEG and 2DHG) and the energy gap $\Delta$. When $\Delta = 0$, the resonant-tunneling transitions with the conservation of momentum can dominate~\cite{1,7,15,16}. Such transitions correspond to
the arrow in Fig.~3(a).
We present the rate of the resonant-tunneling recombination in the following form:

\begin{equation}\label{10}
R_{rt} \simeq  \frac{\Sigma_T}{\tau_{rt}}\exp\biggl(-\frac{\Delta^2}{\gamma^2}\biggr),
\end{equation}
where  $\tau_{rt}$ is  the characteristic  resonant-tunneling recombination time and $\gamma$ is the resonance broadening parameter.

The inter-GL transitions  due to the scattering on impurities or acoustic phonons (non-resonant processes) and due to
the processes mediated  by photons see Fig.~3(b)] in which the momentum or energy are not conserved can
also contribute to the recombination.
Due to a strong coupling of electrons and holes with optical phonons,
the processes assisted by optical phonon emission can also greatly contribute to the rate of the inter-GL transitions (as it takes place in GL structures with bipolar injection into GLs~\cite{10,14,17,18,19}).
For concreteness, we assume that the  resonant-tunneling processes and the processes accompanied by oprical phonon emission are
the main mechanisms of  the inter-GL  recombination.

%
%
%
In the case $V < \hbar\omega_0/e \simeq 200$~mV,
  $2\varepsilon_F + \Delta < \hbar\omega_0$, and the transitions between the tails
  of the energy distributions are possible. At $eV > \hbar\omega_0 \simeq 200$~meV, the inter-GL transitions assisted by the optical phonon emission
are not limited by the Pauli exclusion principle [see Fig.~3(b)]. 
In this case, the inter-GL recombination can be rather strong restricting penetration of the injected electrons and hole sufficiently far from  GL edges. Considering this,  
 the rate of the nonresonant inter-GL recombination assisted by optical phonon emission
 can be presented as 
 
\begin{equation}\label{11}
R_{nr} \simeq 
 \frac{\Sigma_T}{\tau_{nr}}\exp\biggl(\frac{2\varepsilon_F + \Delta}{T}\biggr)
\end{equation}
if $\varepsilon_F + \Delta < \hbar\omega_0$, and

\begin{equation}\label{12}
R_{nr} \simeq 
 \frac{\Sigma_T}{\tau_{nr}}\exp\biggl(\frac{\hbar\omega_0}{T}\biggr)
\end{equation}
if $\varepsilon_F + \Delta > \hbar\omega_0$
Here $\tau_{r} \simeq (\Sigma_T/G_T)\exp(2\ae\,d)$ is the characteristic recombination time associated with  and with the inter-GL transitions assisted by the optical phonon 
emission, respectively,
$G_T$ is the rate thermogeneration of the electron-hole pairs due to the absorption of optical phonons  in equilibrium (it is about $G_T = 10^{21}$~cm$^{-2}$s$^{-1}$ at room temperature~\cite{16}), and $\ae$ is the tunneling decay factor characterizing the overlap of wavefunctions in the top and bottom GLs, 
The right-hand side of 
Eq. (11), which provides a somewhat simplified dependence of the recombination rate on $\varepsilon_F$ and $\Delta$, which differs from that used in Ref.~\cite{14,15} (see also Refs.~\cite{16,17}) for the recombination rate
associated with the intra-GL transitions mediated by optical phonons  by factors $\exp(- 2\ae\,d)$ and $\exp(\Delta/T)$. The latter is due to the energy gap associated with the GL spatial separation and the potential difference between GLs.
Invoking Ref.~\cite{17}, we find $\Sigma_T/G_T \simeq 
(10^{-9} - 10^{-10})$~s. 
Hence, accounting for that $\exp(2\ae\,d) \gg 1$, one obtains $\tau_{nr} \gg
(10^{-9} - 10^{-10})$~s.
According to the experimental results~\cite{1}, the rate of the inter-GL transitions at the resonant tunneling $R_{rt} \simeq 10^{22}$~cm$^{-2}$s$^{-1}$. This yields $\tau_{rt} \simeq (1 - 3)\times 10^{-11}$~s$^{-1}$.

Thus, the net inter-GL recombination rate is assumed to be as

\begin{equation}\label{13}
R =  R_{nr} + R_{nr}.
\end{equation}

Considering the same  boundary conditions as in Ref. \cite{14} set at the points near
the contacts (outside narrow space-charge regions) i.e., at $x = \pm L^* =(L - r_s) \simeq \pm L$,  but generalized by accounting for the energy gap (see Fig.~2), we have
\begin{equation}\label{14}
(e\varphi_h + \varepsilon_{F})|_{x = -L} = \frac{eV}{2},
\qquad (e\varphi_e - \varepsilon_{F})|_{x = L} = -\frac{eV}{2}.
\end{equation}.

The boundary conditions for the electron and hole velocities can be  taken as
(no electron and hole current at the disconnected GL edges)

\begin{equation}\label{15}
u_e|_{x = - L} = u_h|_{x = L} = 0.
\end{equation}.


As in Ref.~\cite{14} , we introduce the following  dimensionless  variables: $\Phi_{e,h} = e\varphi_{e,h}/k_BT$, 
$\mu = \varepsilon_{F}/T$, $\mu_g = \varepsilon_{F_g}/T$, $v = eV/T$, $v_0 = eV_0/T$, $\delta = \Delta/T$,
    $U_e = u_e\tau_{nr}/L$, $U_h = u_h\tau_{nr}/L$, and
$\xi = x/L$.   In these notations, the dimensionless
generation-recombination term $r(\mu) = R\tau_{nr}/\Sigma_T$ and 
the dimensionless density $\sigma(\mu) =  \Sigma/\Sigma_T$ are

\begin{equation}\label{16}
\frac{d[\sigma(\mu)\,U_e]}{d \xi} = -r(\mu),\qquad \frac{d [\sigma(\mu)\, U_h]}{d x} = - r(\mu).
\end{equation}

\begin{equation}\label{17}
\frac{d(\Phi_e - \mu)}{d\xi} = 
\frac{\beta_{eh}(\mu)}{\sigma(\mu)}\biggl[U_e\biggl(\frac{\nu}{\nu_{eh}}\biggr) + U_e - U_h)\biggr],
\end{equation}

\begin{equation}\label{18}
-\frac{d(\Phi_h + \mu)}{d\xi} = \frac{\beta_{eh}(\mu)}{\sigma(\mu)}\biggl[U_h\biggl(\frac{\nu}{\nu_{eh}}\biggr) + U_h - U_e)\biggr],
\end{equation}
\begin{equation}\label{19}
\sigma(\mu) = \frac{12}{\pi^2}\int_0^{\infty}\frac{dyy}{[\exp(y - \mu) +1]}.
\end{equation}

$$
r(\mu) = \exp[2\mu + 2(\mu^2 - \mu_g^2)/v_0] 
$$
\begin{equation}\label{20}
+ \eta\exp[- 4b^2 (\mu^2 - \mu_g^2)^2],
\end{equation}
where  $\beta_{eh}(\mu) = [M(\mu)\nu_{eh}(\mu)\sigma(\mu))L^2/T\tau_r)
= {\overline \beta}_{eh} I(\mu)$, 
where the function 
$I(\mu)$ is numerically calculated  in Appendix,
$v_0 = (\hbar^2v_W^2\kappa/2e^2dT)$,  $b = (T/v_0\gamma)$, and $\eta =\tau_{nr}/\tau_{rt}$.  
The parameter $\beta_{eh}$ varies in a wide range depending on 
on the scattering and recombination parameters and the GL length.

The quantity $q_{eh} = \beta_{eh}/\sigma$ can be presented as $(L/{\cal L}_{eh})^2$, where ${\cal L}_{eh} = D_{eh}\tau_r$
is the diffusion length and $D_{eh} = v_W^2/2\nu_{eh}$ is the bipolar diffusion coefficient.  


\section{Spatial and voltage dependences of  Fermi Energy, energy gap, and potential}

%
In the structures under consideration in which the density of 2DEG and 2DHG markedly exceeds the density residual impurities, one can assume that $\nu \ll \nu_{eh}$ (even despite the spatial separation of 2DEG and 2DHG).
Taking this into account, we disregard in Eqs.~(10) and (11) the terms proportional to a small parameter
$\nu/\nu_{eh}$. 
 In such a case, Eqs.~(16) - (18), with boundary conditions which follow from Eqs.~(14)
 and (15), yield
$\mu = \mu_i = \varepsilon_{Fi}/T = const$, $\delta = \delta_i = const$,  $\Phi_{e} = -\delta_i/2 + \Phi_i$,
and $\Phi_{h} = \delta_i/2 + \Phi_i$, where
$\Phi_{i} \propto \xi$:

\begin{equation}\label{21}
U_e = - \frac{r(\mu_i)}{\sigma(\mu_i)}(\xi + 1), \qquad 
U_h = - \frac{r(\mu_i)}{\sigma(\mu_i)}(\xi - 1),
\end{equation}

\begin{equation}\label{22}
\Phi_i = - \frac{2\beta_{eh}(\mu_i)r(\mu_i)}{\sigma^2(\mu_i)}\,\xi,
\end{equation}

\begin{equation}\label{23}
\delta_i = 2(\mu_i^2 - \mu_g^2)/v_0.
\end{equation}
As follows from the boundary conditions, $\mu_i$ and $\Delta_i$
  are related  to each other also as:

\begin{equation}\label{24}
\mu_i + \frac{2\beta_{eh}(\mu_i)r(\mu_i)}{\sigma^2(\mu_i)} = \frac{v - \delta_i}{2},
\end{equation}
Considering Eqs.~(22) and (23), we arrive at
the following equation for $\mu_i$:

\begin{equation}\label{25}
\mu_i +   \frac{(\mu_i^2 - \mu_g^2)}{v_0} 
+ \frac{2\beta_{eh}(\mu_i)r(\mu_i)}{\sigma^2(\mu_i)} = \frac{v}{2}.
\end{equation}
The quantity $\delta_i$ in  Eq.~(23), which is proportional to $v_0^{-1} \propto d$, explicitly describes 
the effect of spatial separation of GLs resulting in the appearance of the energy gap.
According to the above formulas, the Fermi energy and the energy gap are independent of the coordinate (in the main parts of GLs except near-contact regions), while potential is a linear function of the coordinate {see Eq.~(22)]. This corresponds to  Fig.~2.

If the inter-GL recombination is insignificant, formally setting $r(\mu) = 0$,
from Eq.~(24), we obtain the equilibrium value of the Fermi energy in each GL, and, hence,
using Eq.~(23), the equilibrium value of the energy gap coinciding with Eqs.~(4) and (5).

 Equation~(25) yields the explicit relationship between  the normalized Fermi energy $\mu_i$ in the main portions of GLs  (excepts narrow regions near the contacts) on
parameters $q$ and  $v_0$,  the bias voltage $V$, and the gate voltage $V_g$ [via the dependence $\mu_g(V_g)$].
At relatively weak electron-hole scattering (and the scattering on disorder)and recombination when $\beta  \ll 1$ and 2DEG and 2DHG are close to equilibrium, Eqs. (23) - (25) yield
$\mu_i$ and $\delta_i$ close to $\mu_i^{eq} = \varepsilon_F^{eq}/T$ and $\delta_i^{eq} = \Delta^{eq}/T$
[see Eqs.~(4) and (5)].

Figure~4 shows the dependences of the  normalized energy gap $\delta_i$
and the normalized Fermi energy, $\mu_i$ in the main parts of GLs  calculated using Eqs.~(23) - (25) for the parameter ${\overline \beta}_{eh} $ in the range of  ${\overline \beta}_{eh} = 0.001 - 0.01$ (${\overline \beta}_{eh}\eta = 0.1 - 1.0$).
It was assumed that  $\kappa = 4$, $d = 2$~nm, $\eta = 100$, $T = 300$~K, and $\mu_g = 2$ (i.e., $V_{bi} = 100$~mV), so that 
$v_0 \simeq 10.5$.  $\gamma = 1$~meV, $b \simeq 2.375$.
We used the  functions $\beta_{eh} (\mu)$ (for $\alpha = e^2/\kappa\hbar\,v_W = 0.505$)
and $\sigma(\mu)$ calculated in Appendix. 

Using the data obtained from Ref.~\cite{1} ($\tau_{rt} \sim (1 - 3)\times 10^{-11}$~s and
$L^2 = 0.6~\mu$m$^2$), one can find ${\overline \beta}_{eh} \sim 0.0017 - 0.0051$.
The quantity $\eta = 100$ corresponds to $\tau_{nr} \sim  (1 - 3)\times 10^{-9}$~s (see Sec.~II).
As seen from Fig.~4, at relatively small values of parameter ${\overline \beta}_{eh}$
both $\delta_i$ and $\mu_i$ increase monotonically with increasing the bias voltage
(see curves 1 and 2).
At elevated values of parameter ${\overline \beta}_{eh}$, for example at relatively long GLs (see curves 3 and 4), the voltage dependences of $\delta_i$ and $\mu_i$
are of the S-shape, i.e.,  multi-valued in a certain voltage range corresponding to the tunneling resonance $\delta_i = 0$). The monotonic and multi-valued behavior correspond, in particular, to relatively short and long GLs, respectively (${\overline \beta}_{eh} \propto L^2$)


\section{Role of  scattering on disorder}

Although the scattering of electrons and holes on residual impurities and acoustic phonons
is comparably weak, it can lead to a small but qualitative modification of the
injection characteristics.
Considering the terms with $\nu$ as perturbations, 
one can search for solutions of Eqs.~(15) - (18) in the form:
$\mu = \mu_i + \delta\mu$,  $\delta - \delta_i  =4q_0\mu_i\delta\mu $, and so on.
In particular, we arrive at the following equation for the variation of the Fermi energy:

\begin{equation}\label{26}
\frac{d \delta\mu}{d\xi} = 
\biggl(\frac{\nu}{\nu_{eh}}\biggr)
\frac{\beta_{eh}(\mu_i)r(\mu_i)}
{(1 +2\mu_i/v_0)\sigma^2(\mu_i)}\,\xi.
\end{equation}
Taking into account that $\delta\mu|_{\xi = \pm 1} = 0$,
from Eq.~(26)
we obtain

\begin{equation}\label{27}
\delta\mu = 
\frac{\beta_d(\mu_i)\,r(\mu_i)}{2(1 +2\mu_i/v_0)\sigma^2(\mu_i)}\,(\xi^2 - 1).
\end{equation}
Here $\beta_d(\mu) = 
(\nu/\nu_{eh})\beta_{eh}(\mu)$. Simultaneously, one can find

\begin{equation}\label{28}
\delta - \delta_i = 
\frac{2\beta_d(\mu_i)\,r(\mu_i)}
{v_0(1 +2\mu_i/v_0)\sigma^2(\mu_i)}\,(\xi^2 - 1).
\end{equation}

As follows from Eqs.~(27) and (28), the scattering of electrons and holes on
disorder results in  the spatial distributions of the Fermi energy $\mu$ (i.e., $\varepsilon_F$) and the energy gap $\Delta$ with minima at the center of the structure
($\xi = 0$). The span of these quantities  variations are as follows:

$$
\delta\mu|_{\xi = 0} = - 
\frac{\beta_d(\mu_i)\,r(\mu_i)}{2(1 +2\mu_i/v_0)\sigma^2(\mu_i)},
$$
\begin{equation}\label{29}
= - 
\biggl(\frac{L}{{\cal L}}\biggr)^2
\frac{r(\mu_i)}{2(1 +2\mu_i/v_0)
\sigma(\mu_i)},
\end{equation}

\begin{equation}\label{30}
(\delta - \delta_i)|_{\xi = 0} 
= - 
\biggl(\frac{L}{{\cal L}}\biggr)
\frac{2r(\mu_i)}{v_0(1 +2\mu_i/v_0)\sigma(\mu_i)},
\end{equation}
 where we have introduced   the diffusion  lengths ${\cal L} = (\nu_{eh}/\nu){\cal L}_{eh} = (\beta_{eh}/\beta_d){\cal L}_{eh} \gg {\cal L}_{eh}$ associated with the scattering on disorder.

\begin{figure}[t]
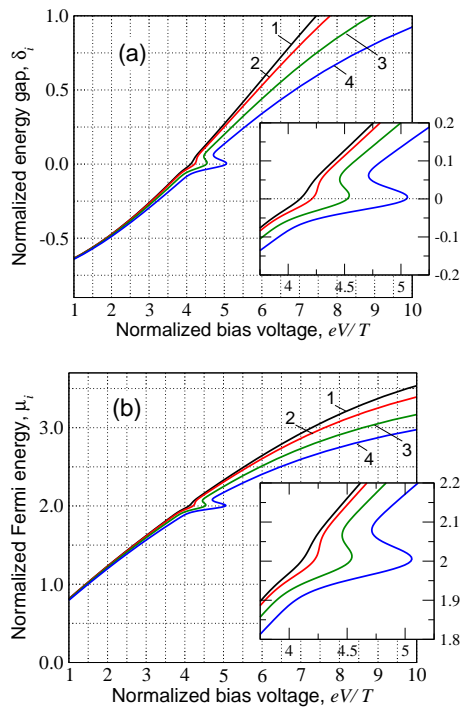

\center{\includegraphics[width=6.0cm]{UND_DGL_DOUBLE_INJECTION_F4a.eps}}\\
\center{\includegraphics[width=6.0cm]{UND_DGL_DOUBLE_INJECTION_F4b.eps}}
\caption{(Color online) Voltage dependences of (a)normalized energy gap between the Dirac points and (b) normalized Fermi energy for different values of parameter ${\overline \beta}_{eh}$: 1 - ${\overline \beta}_{eh} = 0.001$, 2 - 0.002, 3 - 0.005, and 4 - 0.01. Insets show zoom of the same characteristics near the tunneling resonance.} 
\label{Fig4}
\end{figure}
\begin{figure}[t]
\center{\includegraphics[width=6.0cm]{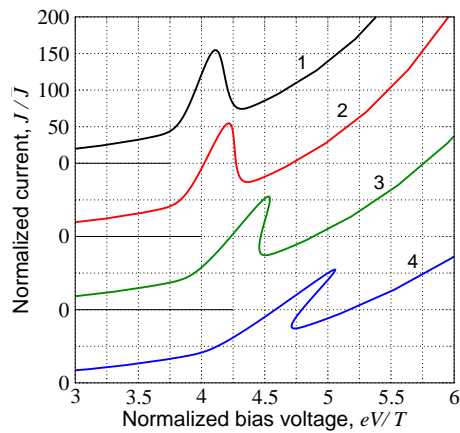}}\\
\caption{(Color online) Current-voltage characteristics for different values of parameter ${\overline \beta}_{eh}$: 1 - ${\overline \beta}_{eh} = 0.001$, 2 - 0.002, 3 - 0.005, and 4 - 0.01; transition from N-shaped to Z-shaped characteristics.} 
\label{Fig}
\end{figure}

\section{Current-voltage characteristics}


The inter-GL current is calculated using the expression for the rate of the pertinent transitions as a function of the local values of the Fermi energy integrated over the GL length:

\begin{equation}\label{31}
J  = \frac{\overline{J}}{2}\int_{-1}^{1}d\xi r(\mu)
\end{equation}
Here 
\begin{equation}\label{32}
\overline{J} = (2e\Sigma_TLH/\tau_{nr}) = (\pi\,eLHT^2/3\hbar^2v_W^2\tau_{nr}),
\end{equation}
$H$ is the width of GLs (in the direction perpendicular to the current), and $r(\mu)$ is given by Eq.~(20).
If $\nu \ll \nu_{eh}$, so that $\mu \simeq \mu_i \simeq const$ in the main parts of GLs,  Eq.~(31) can be rewritten as

\begin{equation}\label{33}
J  \simeq \overline{J}r(\mu_i) = J_i.
\end{equation}
At the tunneling resonance at which $\delta_i = 0$ and $\mu_i = \mu_g$,
the current is equal to (if $2\mu_g < \hbar\omega_0/T$, i.e., $\mu_g < 4$)

\begin{equation}\label{34}
J_i^{max} \simeq {\overline J}\,(e^{2\mu_g} + \eta) = {\overline J}(\displaystyle e^{eV_{bi}/T} + \eta)
\end{equation}
if $eV_{bi} < \hbar\omega_0$, and

\begin{equation}\label{35}
J_i^{max} \simeq {\overline J}(\displaystyle e^{\hbar\omega_0/T} + \eta)
\end{equation}
if $eV_{bi} >  \hbar\omega_0$.
When  $\mu_g = 2$, i.e., $V_{bi} = 100$~mV,  ${\overline \beta}_{eh} = 0.001$, and $\eta = 100$, Eq.~(34) yields 
$J_I^{max}/{\overline J} \simeq 150$. 
%
%
At higher values of $\mu_g \propto V_{bi} \propto \sqrt{V_g}$,  the voltage dependence of   the current peak height 
  tends to saturation  or becomes
relatively slow  if a relatively weak (non-exponential) dependence on the non-resonant inter-GL transition rate is accounted for .

Figure~5 shows the current-voltage characteristics calculated using Eq.~(30) with
Eqs.~(20), (23), and (25) for the same parameters as in Fig.~4.
As seen from Fig.~5, the current-voltage characteristics exhibit pronounced maxima
corresponding to the tunneling resonances. Indeed, the bias voltage
at which the current reaches the maximum coincide with that corresponding to $\delta_i = 0$
[see Fig.~4(a)]. It should be noted that the width, $\Delta V$, of the maxima markedly exceeds
the width, $\gamma/e = 1$~mV, of the resonant  maximum of the inter-GL transition rate  $r(\mu)$ by the order of magnitude.
However, the most remarkable feature of the obtained current-voltage characteristics
is the transformation of their shape from the N-type (curves 1 and 2) to the Z-type (curves 3 and 4) when the parameter ${\overline\beta}_{eh}$  increases. As seen from Fig.~5, the current-voltage characteristics become 
 multi-valued (when the voltage  dependences  of the Fermi energy and the energy gap become of the S-shape) in  certain voltage ranges  if the parameter ${\overline \beta}_{eh}$ is sufficiently large (i.e., if the length of GL $L $ is relatively large).
The effect of broadening of the resonant maxima in the current-voltage characteristics 
is due to relatively slow variations of the energy gap with varying bias voltage. Indeed,
the width of the current-voltage resonant  peak can be estimated as 
$\Delta V/ (\gamma/d\delta_i/dv) = \Delta V/ (\gamma/d\Delta_i/dV)$. 
Since  $d\delta_i/dv  = d\Delta_i/dV \ll 1$ [see Fig.~4(a)], $\Delta V \gg \gamma$. 
Even at ${\overline \beta}_{eh} \ll 1$, when 
2DEG and 2DHG in the pertinent GLs are close to equilibrium,  $d\delta_i/dv \simeq 2\mu_g/(v_0 + 2\mu_g)
= V_{bi}/(V_0 + V_{bi}) \simeq 0.275$ for the parameters used in the above calculations. An increase in $\beta$  results in further decreasing 
of $d\delta_i/dv$ and, hence, increasing the width of the current-voltage resonant  peak.
The  occurrence of the voltage range where they are multi-valued, i.e., the transformation of the N-shaped current-voltage characteristics to the Z-shaped ones can be attributed to
the potential drop across GLs. Similar effect in quantum-well resonant-tunneling transistors
and other tunneling devices  was considered previously~\cite{21,22,23,24,25} (see also Ref.~\cite{26}).
For the values $\tau_r = (1 - 3)\times 10^{-9}$~ s$^{-1}$ and $\tau_{rt} = (1 - 3)\times 10^{-11}$~ s$^{-1}$ used above and $2LH = 0.6~\mu$m$^2$ (as in Ref.~\cite{1}), the tunneling resonant peak current $J_i^{max}$ ($J_I^{max}/{\overline J} \simeq 150$) is in the range  $50 - 150$~nA, i.e., of the same order of magnitude as in the experiment~\cite{1}.

An increase in the gate voltage $V_g$ leads to an increase in the Fermi energy $\varepsilon_F$ (i.e., $\mu$) and, hence, to  an increase in $V_{bi}$ (i.e., $\mu_g$). As a result, the position of the current peak shifts to
higher bias voltages. 


The shift of the current maxima associated with an increase   in
$\varepsilon_F$ and, consequently, $V_{bi}$  
occurs when the levels of the chemical doping  of GLs  are elevated 
(the donor density in the  top GL  and the acceptor density in the bottom GL increase). In this case,  the resonant condition is achieved at higher bias voltage $V$.  Just such doped structures were studied in Ref.~\cite{1}, in which the current peaks correspond to higher values of the bias voltages ($V$ about several tenth of Volt) than in 
Fig.~5 ($V \sim 0.1$~V) for the chemically undoped GLs.  However, an increase in the dopant densities leads to
an increase in the collision frequency $\nu$. This can result in a marked effect of the disorder scattering
to the spatial distributions of the Fermi energy, the energy gap [see Eqs.~(27) and (28)],  and  the current.
 Due to this, the function $r(\mu)$ in Eq.~(31)
becomes coordinate-dependent. 
This  leads to lowering of the current peak and  its additional  broadening (inhomogenious).
Indeed,
using Eqs.~(27) and (31) at $\mu_i = \mu_g$ (when $\delta_i = 0)$ and integrating over $d\xi$, we obtain

\begin{equation}\label{36}
J^{max} \simeq J_i^{max}\biggl[1   + \frac{\beta^2_d(\mu_g)r^2(\mu_g)}{15(1 + 2\mu_g/v_0
)^2\sigma^4(\mu_g)}r^{\prime\prime}(\mu_g)\biggr], 
\end{equation}
where $r^{\prime\prime} (\mu_g) = d^2r(\mu_i)/d\mu_g^2$. 
Since $r^{\prime\prime} (\mu_g) \simeq - 12b^2\mu_g^2$ is negative, from Eq.~(33) we obtain
$J^{max} <  J_i^{max}$ with

$$
 J_i^{max} - J^{max} \simeq \frac{12b^2\beta^2_d(\mu_g)(e^{2\mu_g} + \eta)^2\mu_g^2}{15(1 + 2\mu_g/v_0
)^2\sigma^4(\mu_g)}
$$

Taking into account that at $\mu > 1$ $\beta_d \simeq {\overline \beta}_d\mu^2$ 
(where ${\overline \beta}_d$ is independent of $\mu_g$a and $\sigma(\mu) \propto \mu^2$, Eq.~(35) can be presented as

\begin{equation}\label{37}
J_i^{max} - J^{max} \sim   
{\overline \beta}^2_{d}b^2\eta^2. 
\end{equation}
The latter quantity is proportional to a small factor ${\overline \beta}^2_{d}$
and two large factors $b^2$ and $\eta^2$. Hence even relatively weak scattering on disorder can markedly affect the shape of the maxima in the current-voltage characteristics.






\section{Conclusions}

In conclusion:

(1) We have developed the device model for the
double-GL structures with  independently contacted GLs which describes the processes of the electron and hole injection from the opposite contacts accompanied with the inter-GL electron-hole recombination.

(2) Using the model,  we have derived analytical expressions for  the spatial
distributions of the electron and hole  Fermi energies and the energy gap between the Dirac points in GLs  as functions of the bias and gate voltages and the structural parameters. It has been shown that these quantities   can be virtually 
coordinate-independent in the main parts of GLs (except narrow near contact regions)  if the  mutual scattering of electrons and holes dominate.
An increase in relative strength of electron and hole scattering on disorder can  lead
to   substantial sag of the spatial dependences in question with the minima in the center of the GL structures.
The shape of the Fermi energy versus voltage characteristics varies from the monotonic to the S-type with increasing length of GLs.

(3)  We have calculated the current-voltage characteristics and revealed that they exhibit maxima associated with
the resonant-tunneling inter-GL transitions (the N-type characteristics) in the GL-structures with relatively short
GLs. In the structures with long GLs, the N-type characteristics can transform to the Z-type characteristics.
In the latter case, the effect of hysteresis can exist.  This should be taken into account choosing the range of operation voltages of  the lasers based on the double-GL structures~\cite{11,27}.

(4) The obtained results are in line with recent experimental observations~\cite{1}. They can be useful for the development and optimization of double-GL-based field-effect transistors, terahertz detectors, and terahertz lasers.

\vspace*{-3mm}

\section*{Acknowledgments}
The authors are grateful to A. Satou and D.~Svintsov   for useful discussions.
This work was  supported  
by the Japan Society for Promotion of Science (Grant-in-Aid for Specially Promoting Research), Japan. The work at UB was supported by the NSF-TERANO Program.

\section*{Appendix}
\setcounter{equation}{0}
\renewcommand{\theequation} {A\arabic{equation}}
\begin{figure}[h]
\center{\includegraphics[width=6.0cm]{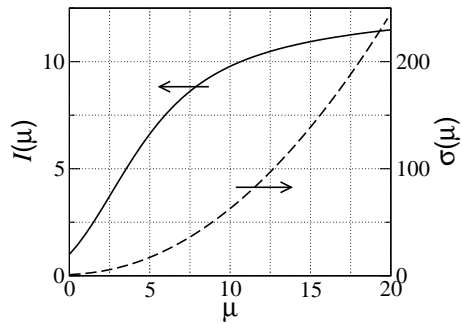}}\\
\caption{Electron-hole collision function $I(\mu)$  and normalized density $\sigma(\mu)$}. 
\label{Fig6}
\end{figure}
The function $\sigma(\mu)$ is given by Eq.~(19). At $\mu \rightarrow 0$, $\sigma(\mu) \rightarrow 1$, while at large $\mu$, one obtains $\sigma(\mu) \simeq 6\mu^2/\pi^2$. 
The function $\beta_{eh} \propto I(\mu) = i(\mu)/i(0)$, where
according to Ref.~\cite{13} (see, also, Ref.~\cite{20}) and considering the Thomas-Fermi screening of the electron-hole interaction,

$$
i(\mu) = \int_0^{\infty}\int_0^{\infty}dxdx^{\prime} F(x - \mu)F(x^{\prime} - \mu)
$$
$$
\times\int_{-x^{\prime}}^xdQ[1 - F(x - Q - \mu)][1 - F(x^{\prime} + Q - \mu)]
$$
\begin{equation}\label{A1}
\times\frac{\sqrt{xx^{\prime}(x - Q)(x^{\prime} + Q)}\,Q^2}{[|Q|) -4\alpha\ln F^2(\mu)]^2},
\end{equation}
where $F(x) = (1 + e^x)^{-1}$ and $\alpha = e^2/\kappa\hbar\,v_W$ (at $\kappa = 4$, $\alpha \simeq 0.505$).
Figure~6 shows $\sigma(\mu)$ and $I(\mu)$ calculated numerically using Eqs.~(16) and (A1), respectively.

\end{document}